\documentclass[aps,showpacs,twocolumn,twoside]{revtex4}
\usepackage{epsfig,amssymb,amsmath}

\begin{document}

\def\ket#1{|#1\rangle}
\def\bra#1{\langle#1|}
\def\scal#1#2{\langle#1|#2\rangle}
\def\matr#1#2#3{\langle#1|#2|#3\rangle}
\def\bino#1#2{\left(\begin{array}{c}#1\\#2\end{array}\right)}
\def\ave#1{\langle #1\rangle}
\def\dis#1{\langle\langle #1\rangle\rangle}
\def\uvo#1{\lq #1\rq\ }
\def\uuvo#1{\lq\lq #1\rq\rq\ }

\title{Impact of Quantum Phase Transitions on Excited Level Dynamics}
\author{Pavel Cejnar, Pavel Str{\' a}nsk{\' y}}
\affiliation{Institute of Particle and Nuclear Physics, Faculty of Mathematics and Physics, Charles University,
V Hole{\v s}ovi{\v c}k{\'a}ch 2, 180\,00 Prague, Czech Republic}

\date{\today}
\begin{abstract}
The influence of quantum phase transitions on the evolution of excited levels in the critical parameter region is discussed. 
The analysis is performed for 1D and 2D systems with first- and second-order ground-state transitions.
Examples include the cusp and nuclear collective Hamiltonians.
Applications in systems of higher dimensions are possible.
\end{abstract}

\maketitle

\section{Introduction}

In the last decade, the concept of quantum phase transitions (QPT) \cite{Hertz76,Gilmore78,Gilmore79} has triggered a lot of activity in solid-state physics, see e.g. Refs.\,\cite{Sachdev99,Vojta03,Belitz05}, and in many-body (nuclear) physics, see e.g. Ref.\,\cite{Cejnar08} and the references therein.
The QPT shows up in the system's infinite-size limit at zero temperature as a nonanalytic change of the ground-state properties with an external control parameter (e.g. an interaction strength).
Examples range from order-disorder transitions in spin lattice systems \cite{Sachdev99} to transitions between various quasidynamical symmetries in some interacting boson/fermion models \cite{Dieperink80,Zhang87,Volya03,Rowe04,Emary03}.

Phase transitional effects were recently identified also for excited states in the two-level pairing models that exhibit a second-order ground-state QPT \cite{Heinze06,Cejnar06,Cejnar07,Caprio08}.
Two characteristic signatures of excited-state transitions were recognized, namely (a) an anomalous evolution of individual excited levels as they cross the phase separatrix in the plane of the control parameter versus energy, and (b) a singular behavior of the level density at the critical energy \cite{Caprio08}. 
In this article, a generalization of these results to systems with the {\em first-order\/} ground-state QPT is discussed.
It is shown that a phase transition at zero temperature has significant consequences for the dynamics and density of excited levels in the critical region of the control parameter.
These features are closely related to thermodynamical properties of the system.

The plan of the paper is the following: 
In Sec.~\ref{led} we introduce a general framework for studying excited-state phase transitions in quantum systems with coinciding classical and thermodynamical limits. 
This type of systems underlies a wide class of models used in many-body physics. 
In Sec.~\ref{cusph}, the concept of excited-state QPT is illustrated by a simple, prototypal example of a one-dimensional model with both first- and second-order ground-state transitions (the cusp catastrophe). 
The dependence of phase-transitional effects on the dimension is discussed in Sec.~\ref{collh}, where an example of a two-dimensional system, closely related to the nuclear collective model, is analyzed. 
The last section contains a brief summary and conclusions.

\section{Quantum phase transitions for excited states}
\label{led}

As mentioned above, quantum phase transitions are related to the zero-temperature limit of the system, thus to properties of the ground state.
An extension of the QPT analysis to excited states can be achieved through the free energy 
\begin{equation}
F=\underbrace{{\rm Tr}(\hat{\varrho}\hat{H})}_{\ave{E}}
+T\underbrace{{\rm Tr}(\hat{\varrho}\ln\hat{\varrho})}_{-S}=-T\ln Z
\,,
\label{free}
\end{equation}
where ${\hat H}$ is the Hamiltonian, $\hat{\varrho}$ the canonical density operator, $Z$ the partition function, $\ave{E}$ an  average energy at temperature $T$, and $S$ the corresponding entropy \cite{Reichl}.
For $T=0$, the free energy coincides with the ground-state energy.

The known relations
\begin{equation}
\frac{\partial F}{\partial T}=-S\,,\quad
\frac{\partial^2 F}{\partial T^2}=-\frac{\dis{E^2}}{T^3}
\,,
\label{freedT}
\end{equation}
where $\dis{E^2}$ stands for a thermal dispersion of energy, set up the conditions for thermal phase transitions: 
While in the first-order transition the entropy suddenly jumps at a certain \lq\lq critical\rq\rq\ temperature $T_{\rm c}$, implying a discontinuous first derivative of $F$, in a continuous phase transition only the second and/or higher derivatives are affected.
Such situations can only occur in the limit of infinite system size, when the canonical description is assumed to converge to the microcanonical one.
Then the entropy can be written as $S\propto\ln\rho$, where $\rho$ is the level density at energy $E=\ave{E}$.

If the Hamiltonian depends on an external control parameter $\lambda$, so does the free energy (\ref{free}).
An interesting question is whether a quantum phase transition at $T=0$ and $\lambda=\lambda_{\rm c}(0)$ extends to $T>0$ in the above thermodynamical sense, forming a phase separatrix $\lambda_{\rm c}(T)$.
Although specific examples exist, in which the $T=0$ critical point is isolated, in generic situations the phase transition exists in the $\lambda\times T$ plane \cite{Hertz76,Gilmore78,Gilmore79,Sachdev99,Vojta03}.
The classification of such transitions is again in terms of the behavior of the free energy at the transitional point, but with respect to both variables $T$ and $\lambda$.
In particular, the first-order or second-order transitions, respectively, imply a discontinuity of the gradient vector or of the curvature matrix (Hessian) associated with $F(\lambda,T)$.

Assume a linear dependence of the Hamiltonian
\begin{equation}
\hat{H}(\lambda)=\hat{H}_0+\lambda\hat{H}_1
\end{equation}
on a single real parameter $\lambda$, with $\hat{H}_0$ and $\hat{H}_1$ standing for two incompatible dynamical terms.
From the perturbation theory it follows that the additional relations to (\ref{freedT}) read as:
\begin{equation}
\frac{\partial F}{\partial\lambda}=\ave{\dot{E}}
\,,\quad
\frac{\partial^2 F}{\partial\lambda^2}=\ave{\ddot{E}}-\frac{\dis{\dot{E}^2}}{T}
\,,\quad
\frac{\partial^2 F}{\partial T\partial\lambda}=\frac{\dis{E\dot{E}}}{T^2}
\,.
\label{freedl}
\end{equation}
Here $\dis{XY}=\ave{XY}-\ave{X}\ave{Y}$, with $\ave{\bullet}$ standing for the thermal average of individual level energies $E_i$ and their derivatives $\dot{E_i}\equiv \tfrac{d}{d\lambda}E_i$, $\ddot{E_i}\equiv\tfrac{d^2}{d\lambda^2}E_i$.

The formulas in Eq.~(\ref{freedl}) can be used to anticipate the implications of thermal phase transitions on the level dynamics.
The first-order transition, which shows up as a discontinuity of $\tfrac{\partial}{\partial T}F$ and $\tfrac{\partial}{\partial\lambda}F$, leads to a jump of the average slope $\ave{\dot{E}}$ of energy levels.
If the phase separatrix is not parallel with $\lambda$ or $T$, the jump of $\ave{\dot{E}}$ is connected with a jump of $\rho$, thus also a jump of the microcanonical entropy $S$.
This may be viewed from an analogy with the ray refraction on a tilted interface---the vertical distance of rays (alias level spacing $\rho^{-1}$) changes at the interface.
On the other hand, the second-order transition, with discontinuous $\tfrac{\partial^2}{\partial T^2}F$ and  $\tfrac{\partial^2}{\partial\lambda^2}F$, is linked to changes of the average level curvature $\ave{\ddot{E}}$ and/or of the slope dispersion $\dis{\dot{E}^2}$.
Singular derivatives generate continuous phase transitions with no Ehrenfest classification.

In the following, we will illustrate the above general findings by concrete examples.
We focus on systems with a finite number of quantum degrees of freedom $f$.
In the many-body context, such systems usually arise from boson/fermion models based on the dynamical Lie algebras of finite dimensions \cite{Iachello06}.
A typical example is a system consisting of a conserved number $N$ of interacting bosons with the single-particle Hilbert space of dimension $n$: in this case $f=n-1$, with $U(n)$ being the dynamical algebra \cite{Frank94}.

A significant feature of the above type of systems is the fact that the infinite-size limit coincides with the classical limit \cite{Gilmore78,Gilmore79,Cejnar08,Cejnar07b}.
In bosonic systems, e.g., the value of $N^{-1}$ can be associated with the Planck constant $\hbar$.
Therefore, the asymptotic-size limit of the level density $\rho(E)$, which constitutes various kinds of excited-state phase transitions, is proportional to the classical phase-space volume ${\cal V}(E)$ available at energy $E$,
\begin{equation}
{\cal V}(E)\!=\!\!\int\!\delta(E-H)\,d^f\!p\,d^f\!x=\frac{d}{dE}\!\underbrace{\int\!\theta(E-H)\,d^f\!p\,d^f\!x}_{{\cal W}(E)}
\,.
\label{volum}
\end{equation}
Here, $H$ stands for the classical Hamiltonian, $\theta$ is the step function, and ${\cal W}(E)$ represents the phase space volume available at energies less than (or equal to) the value $E$.
This formula will be employed to classify the excited-state QPTs in the forthcoming examples.

\section{Cusp Hamiltonian}
\label{cusph}

Consider first a one-dimensional Schr{\"o}dinger equation
\begin{equation}
\hat{H}=-\frac{K^2}{2}\frac{d^2}{dx^2}+x^4+ax^2+bx
\,,\label{cusp}\end{equation}
with a potential having the well known cusp form \cite{Stewart82}.
Here, $a$ and $b$ are the cusp parameters and $K=\frac{\hbar}{\sqrt{M}}$ is a classicality constant bearing information on the Planck constant and on the mass parameter $M$.
As discussed above, in the bosonic models $\hbar$ is inversely proportional to the size of the system, measured by the boson number $N$, the effective mass depending on a concrete application.
Note that the Hamiltonian (\ref{cusp}) can be transformed \cite{Cejnar07b} to a form equivalent to the model containing two types of interacting (pseudo)scalar bosons (the bosonic formulation of the Lipkin model \cite{Lipkin65,Vidal06}), although in the latter case the kinetic part is generally more complicated than that considered here \cite{Caprio08}.
While the difference in kinetic terms obscures a direct comparison of $K$ and $N$, it is clear than the limit $K\to 0$ corresponds to $N\to\infty$. 

The cusp potential (with germ $x^4$) represents the most common type of catastrophe \cite{Stewart82} in dimension one which generates both first-order and second-order phase transitions.
These can be associated with two trajectories in the plane $a\times b$, namely with potentials $V_1=x^4-x^2+\lambda x$ and $V_2=x^4+\lambda x^2$ depending on a single parameter $\lambda$.
The potential $V_1$ has two minima at $x\neq 0$ within the region demarcated by a pair of spinodal points at $\lambda=\pm\tfrac{4}{3\sqrt{6}}$.
If $\lambda$ in $V_1$ varies from negative to positive values, the ground state exhibits a first-order phase transition---the swap of both minima at the \lq\lq critical\rq\rq\ point $\lambda_{\rm c}=0$.
On the other hand, if $\lambda$ in $V_2$ varies from positive to negative values, the ground state shows a second-order phase transition---the single minimum at $x=0$ ($\lambda>0$) splits at $\lambda_{\rm c}=0$ into a pair of degenerate $x\neq 0$ minima ($\lambda<0$).

It is clear that the cusp potential generates the above types of ground-state QPTs only in the semi-classical limit $K\to 0$ (equivalent to $N\to\infty$ in the bosonic formulation), when the zero point motion vanishes.
The question is what happens in this limit with excited states close to the critical point.
Let us first analyze the behavior of the phase-space volume (\ref{volum}).
It turns out that ${\cal V}(E)$ has some non-analytic features, which are connected with the bottom energies $E^{(1)}$ and $E^{(2)}$ of both potential wells and with the top energy $E^{(3)}$ of the barrier in between.
For a particle moving in a parabolic minimum with classical frequency $\omega$ we find ${\cal V}=\tfrac{2\pi}{\omega}$. 
Therefore, for the double well potential (both minima are locally quadratic) the phase-space volume ${\cal V}(E)$ starts from a non-zero value at $E=E^{(1)}$ and jumps to a higher value at the energy of the secondary minimum $E^{(2)}$, where new states become available in the second well. 
This results in a {\em first-order\/} phase transition.

\begin{figure}
\begin{center}
\epsfig{file=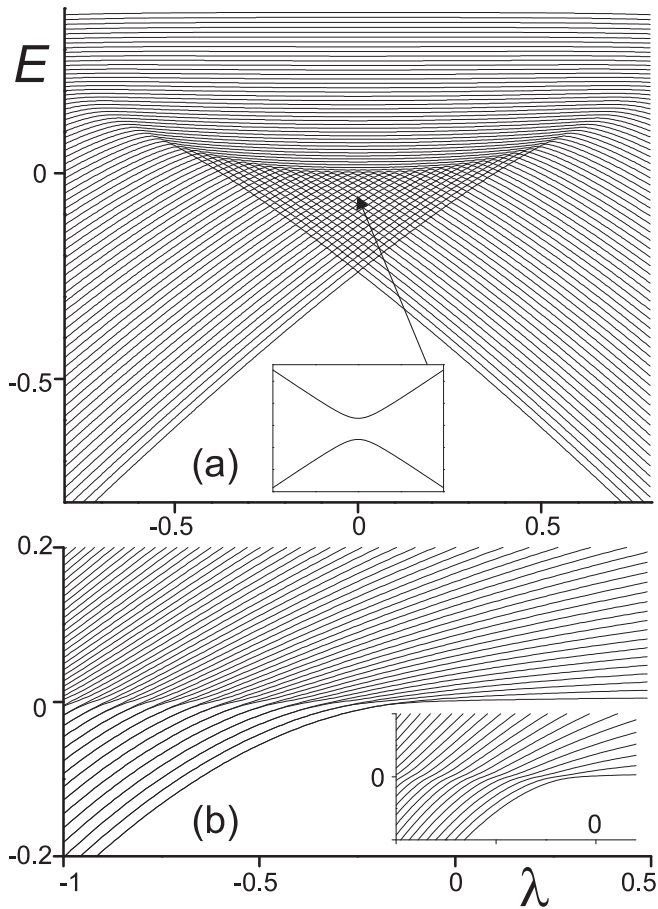,width=\linewidth}
\end{center}
\caption{\protect\small Level dynamics in (a) first-order and (b) second-order phase transition generated by the cusp Hamiltonian (\ref{cusp}) with $(a,b)=(-1,\lambda)$ and $(\lambda,0)$, respectively; $K=10^{-2}$. The inset in panel (b) shows the subset of levels with positive parity.}
\label{levcusp}
\end{figure}

The excited level dynamics for the cusp potential $V_1$ is shown in panel (a) of Fig.\,\ref{levcusp}.
Note that the numerical diagonalization was performed in a truncated oscillator basis, the convergence issues being fully under control. 
The $\nabla$-shaped region, which is apparent in the figure, coincides with the $\lambda\times E$ domain where states can be located in both potential wells.
As levels hit this domain, their \lq\lq laminar\rq\rq\ flow turns into a \lq\lq turbulent\rq\rq\ one.
Note that although the mutual approach of levels in the $\nabla$-region is rather sharp for low-lying states, the crossings are always avoided due to the tunneling effects (see the inset).
As the slopes of up- and down-going stretches of an individual level trajectory compensate each other, the average slope changes abruptly in transition to the $\nabla$-region.
Indeed, this leads to a sudden increase of the level density, as anticipated in the previous discussion.

\begin{figure}
\begin{center}
\epsfig{file=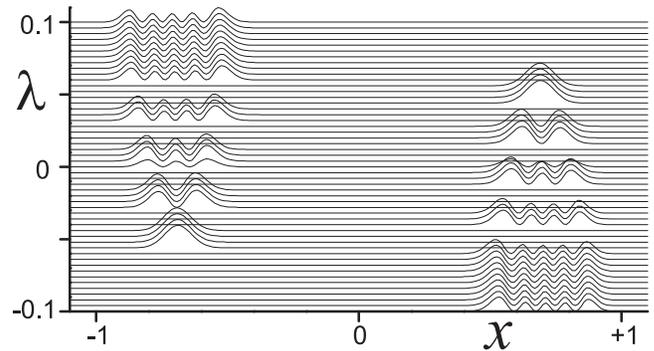,width=\linewidth}
\end{center}
\caption{\protect\small Squared wave function of the 5th state from Fig.\,\ref{levcusp}(a). The $|\psi|^2$ axis is not drawn.}
\label{wavcusp}
\end{figure}

The zig-zag pattern of level trajectories developed in the $\nabla$-region, Fig.\,\ref{levcusp}(a), is connected with two families of states localized in the first and second potential well.
Under the neglect of tunnelling, these states form approximate eigenstates of the Hamiltonian.
While states in the up-going well rise in energy, those in the down-going well decline.
As an eigenstate propagates through this region, its wave function alternates between the two localizations, with the changes taking place at each avoided crossing.
This is illustrated in Fig.\,\ref{wavcusp}, where one observes that the structures with 1, 2, 3 etc. peaks cross the gaps between neighboring levels and continue along the direction in which the respective well moves.
Interestingly, the wiggling pattern of energy levels inside the $\nabla$-region becomes infinitely dense (undifferentiable) in the semiclassical limit.

On the top $E=E^{(3)}$ of the quadratic barrier separating both minima of the cusp potential the phase-space volume ${\cal V}(E)$ exhibits a logarithmic singularity.
It is connected with asymptotic dwell times of classical trajectories at the maximum which yield a locally vanishing gap between quantum levels.
This singularity was analyzed e.g. in Ref.\,\cite{Cary92}.
Its impact on the level dynamics is observed at the upper side of the $\nabla$-region in Fig.~\ref{levcusp}(a).
When level $i$ reaches the energy $E^{(3)}$, the curvature $\ddot{E}_i$ becomes discontinuous and infinite for $K\to 0$.
This leads to a locally infinite growth of the level density and implies a kind of continuous phase transition (with no order).
Let us note that the decrease of $\rho(E)$ after the singularity is connected with a concave increase of ${\cal W}(E)$, see Eq.~(\ref{volum}), which indicates anomalous thermodynamical properties \cite{Reichl}.
As shown below, this peculiarity is specific for 1D (or quasi-1D) cases.

Panel (b) of Fig.\,\ref{levcusp} shows the level dynamics for the second-order phase-transitional potential $V_2$.
The behavior observed is essentially the same as that in the Lipkin model without parity violating interactions.
All levels are characterized by the parity quantum number and for $E\ll 0$ they form approximately degenerate parity doublets.
The phase-transitional evolution is detected at energy $E=E^{(3)}=0$ corresponding to the local maximum of the potential at $x=0$.
This transition has been studied in the Lipkin model \cite{Leyvraz05} as well as in a wider class of two-level pairing models used in nuclear and molecular physics \cite{Cejnar06,Cejnar07,Caprio08,Curro08}.
Let us note that the thermodynamical description of this effect, based on the mean-field approach, was presented in Ref.\,\cite{Gilmore78}.

\section{2D collective Hamiltonian}
\label{collh}

The cusp catastrophe applies to ground-state QPTs in a large class of models, including e.g. two-level interacting boson models \cite{Cejnar07b}.
However, since the number of degrees of freedom $f$ is usually larger than 1, the excited spectrum differs from that in the cusp example.

For $f>1$, only a fraction of trajectories may cross the stationary point, hence a general trend is a smoothening of non-analytic features at the critical energy.
In particular, for $f=2$ only $\tfrac{d}{dE}{\cal V}$ evolves in a discontinuous way when crossing the energy of a local minimum or maximum of the potential, i.e. the transition is of the second order in both cases. 
For a saddle point, the derivative $\tfrac{d}{dE}{\cal V}$ is \lq\lq continuous\rq\rq\ but infinite (${\cal V}$ has a singular tangent).
A further increase of $f$ leads to even higher orders of transitions.
These statements can be verified by explicit evaluation of Eq.~(\ref{volum}) for the respective forms of the potential in general dimension.
As a consequence, for $f>1$ one obtains {\em continuous\/} (second-order or softer) phase transitions along all three sides of the $\nabla$-shaped region of phase coexistence in the first-order QPT.

To illustrate these conclusions, we analyze the following $f=2$ Hamiltonian
\begin{equation}
\hat{H}=-\frac{K^2}{2}\left[\frac{1}{r}\frac{\partial}{\partial r}r\frac{\partial}{\partial r}+
\frac{1}{r^2}\frac{\partial^2}{\partial\varphi^2}\right]+r^4+Ar^2+Br^3\cos 3\varphi
\,,
\label{geo}
\end{equation}
where $(r,\varphi )$ are polar coordinates and $\{K,A,B\}$ adjustable parameters ($K$ has the same meaning as in the cusp case).
We consider only the states with a periodic boundary condition on the sextant $\varphi\in[0,\tfrac{\pi}{3}]$.
This system is closely related to the geometric model and the interacting boson model of collective motions in atomic nuclei \cite{Frank94,Bohr98}.
These models describe 5 degrees of freedom related to nuclear quadrupole deformations, i.e. 2 deformation parameters $(\beta,\gamma)\equiv(r,\varphi)$ and 3 Euler angles, but for states with zero spin the Euler angles become frozen.
The potential reads as that in Eq.~(\ref{geo}).
Let us note the genuine 5D nuclear Hamiltonian for zero-spin states has a slightly different kinetic term than the 2D Hamiltonian (\ref{geo}), but both excitation spectra show the same qualitative features \cite{tobe}.

\begin{figure}[t]
\begin{center}
\epsfig{file=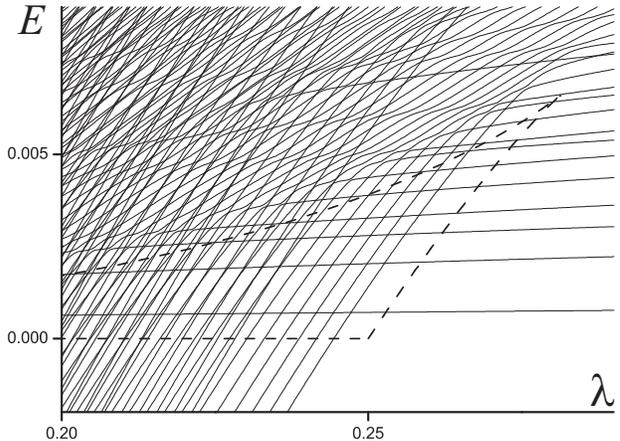,width=\linewidth}
\end{center}
\caption{\protect\small Level dynamics for Hamiltonian (\ref{geo}) close to the first-order phase transition ($A\equiv\lambda$, $B=1$, $K=10^{-3}$). Dashed lines demarcate the phase-coexistence domain.}
\label{ledynG}
\end{figure}

For $K\to 0$, the Hamiltonian (\ref{geo}) exhibits a QPT along the parabola $B^2=4A$ that separates two basic forms of the potential.
The transition is of the first order except at $B=0$ where it is of second order.
On the inner side of the parabola, the potential describes a quartic oscillator with the global $E=0$ minimum at $r=0$.
On the outer side, the potential with $\varphi\in[0,\tfrac{2\pi}{3})$ has the global $E<0$ minimum at $r>0$, $\varphi=\tfrac{\pi}{3}$ (for $B>0$) and a saddle point (with $E<0$) at $r>0$, $\varphi=0$.
In the spinodal region of the first-order transition, the $r>0$ minimum coexists with the $r=0$ one, creating another saddle point (with $E>0$) in between (at $\varphi=\tfrac{\pi}{3}$).
For $A<0$, the $E=0$ minimum at $r=0$ turns into a local maximum.

We consider a scaled potential $V=r^4+\lambda r^2+r^3\cos 3\varphi$ with the critical point $\lambda_{\rm c}=\tfrac{1}{4}$ and two spinodal points at $\lambda=0$ and $\tfrac{9}{32}$.
The level dynamics in the phase-transitional region, obtained by a numerical diagonalization of the Hamiltonian in a modified 2D oscillator basis, is depicted in Fig.\,\ref{ledynG}.
Computational details will be given elsewhere \cite{tobe}.
As in Fig.~\ref{levcusp}(a), all levels pass through virtual (avoided) crossings.
The legs of the phase-coexistence \lq\lq triangle\rq\rq\ (dashed lines) correspond to energy $E^{(2)}$ of the secondary potential minimum (a second-order phase transition) and the upper side $E^{(3)}$ represents a saddle point (a continuous phase transition with a singular tangent of ${\cal V}$).

Despite some similarities, the picture is less dramatic than that for the 1D cusp Hamiltonian.
Under a scrutiny one may observe that the refraction of levels along the boundaries changes the dispersion of slopes rather than their average, consistently with the softer types of phase transitions.
Note that for $B=0$ one can separate subsets of levels, characterized by the 2D angular momentum quantum number $m$, with gradually lowering phase-transitional signatures \cite{Cejnar06,Caprio08}, but this cannot be done in general as $m$ is not conserved for $B\neq 0$.

\begin{figure}[t]
\begin{center}
\epsfig{file=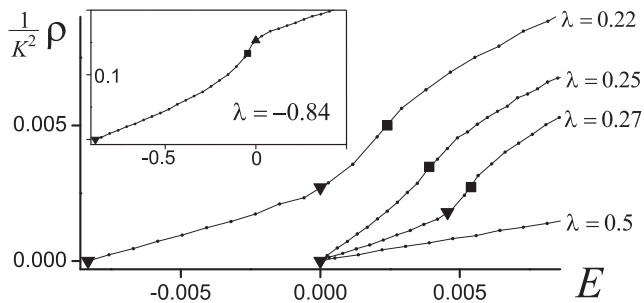,width=\linewidth}
\end{center}
\caption{\protect\small The level density for Hamiltonian (\ref{geo}) with $(A,B)=(\lambda,1)$. Parameter $K$ was set separately for each histogram within the bounds $10^{-4}\leq K\leq 5\cdot 10^{-3}$. The binning of histograms is shown by small dots. Marks indicate minima ($\blacktriangledown$), maxima ($\blacktriangle$) and saddle points ({\tiny$\blacksquare$}) of the corresponding potentials.
}
\label{leden}
\end{figure}

We have evaluated the level density from the numerical spectra at selected values of $\lambda$ and show the result in Fig.~\ref{leden}.
Parameter $K$, which scales the absolute density of states, was chosen for each $\lambda$ to ensure a sufficiently dense spectrum.
While the non-degenerate double-well potentials ($\lambda=0.22$, $0.27$) generate typical forms with three characteristic slopes of $\rho$ in the regions $E<E^{(2)}$, $E^{(2)}<E<E^{(3)}$, $E^{(3)}<E$ (see the marks), the critical potential ($\lambda=0.25$) yields a shape with only one transitional point.
Changes of $\rho$ at the critical energies become sharp in the $K\to 0$ limit.
The level density above the coexistence region ($\lambda=0.5$) exhibits no transitional features.
In contrast, the spectrum below the lower spinodal point (see the inset) displays transitions connected with the $\varphi=0$ saddle point and with the $r=0$ local maximum.

Apart from indicating the phase structure of the specific model under study, the curves in Fig.~\ref{leden} illustrate the detectability limits for continuous phase transitions in a finite system.
Let us stress that the transitions would be even more smoothened in higher dimensions.
An application of these results in the interacting boson model (for the zero-spin spectrum) is in order.

\section{Conclusions}
\label{conclu}

We have investigated the influence of phase transitions on the level dynamics in a vicinity of quantum critical points.
Some typical QPT-induced effects have been identified, depending on the type of the transition and on the dimension.
Of a particular interest are the results for first-order ground-state transitions which supplement earlier findings for the second-order transitions \cite{Cejnar06,Cejnar07,Caprio08}.
In dimension one, the cusp Hamiltonian (\ref{cusp}) is a fundamental example of both first-order and continuous transitions.
Despite the fact that the relevant physics in this case is just the basic-level quantum mechanics, the effects observed constitute the clearest realization of quantum phase transitions affecting individual excited states.

For higher dimensions, the signatures of excited-state phase transitions are weakened in a twofold sense: (a) The transitions only affect higher derivatives of the level density, and (b) only some bulk properties of the level dynamics are influenced rather than evolutions of all individual states. 
These features---which on the classical level can be linked to an increasing size of the phase space, hence a decreasing impact of singular (e.g. stationary) points on the classical motions---hinder the practical detection of excited-state QPTs in finite samples of the system.

Finally, let us stress that the findings discussed in this article are relevant for the systems with synonymous infinite-size and classical limits.
This feature allows one to associate the asymptotic level density with the phase-space volume and therefore to unambiguously classify the excited-state phase transitions.
Numerous examples of this type of systems can be found in nuclear, molecular, and mesoscopic physics.
It would be interesting to learn how the features discussed here extend to the infinite lattice systems that do not share the above property.\\

\acknowledgments

Discussions with F. Iachello and M. Caprio are gratefully acknowledged.
This work was supported by the Czech Science Foundation (202/06/0363) and by the Czech Ministry of Education, Youths and Sports (MSM 0021620859 and LA 314).

\def\Journal#1#2#3#4{{#1} {#2}, {#3} (#4)}
\def\ANNP{Ann. Phys. (N.Y.)}
\def\MATH{J. Math. Phys.}
\def\JPA{J. Phys. A: Math. Theor.} 
\def\JPAold{J. Phys. A: Math. Gen.} 
\def\PRL{Phys. Rev. Lett.}
\def\PREP{Phys. Rep.}
\def\PRA{Phys. Rev. A}
\def\PRC{Phys. Rev. C}
\def\PRB{Phys. Rev. B}
\def\PRE{Phys. Rev. E}
\def\RMP{Rev. Mod. Phys.}
\def\RPP{Rep. Prog. Phys.}
\def\PLB{Phys. Lett. B}
\def\NPA{Nucl. Phys. A}
\def\PPNP{Prog. Part. Nucl. Phys.}


\thebibliography{99}
\bibitem{Hertz76} J. Hertz, \Journal{\PRB}{14}{1165}{1976}.
\bibitem{Gilmore78} R. Gilmore, D.H. Feng, \Journal{\NPA}{301}{189}{1978}.
\bibitem{Gilmore79} R.~Gilmore, \Journal{\MATH}{20}{891}{1979}.
\bibitem{Sachdev99} S. Sachdev, {\em Quantum Phase Transitions} (Cambridge University Press, Cambridge, 1999). 
\bibitem{Vojta03} M. Vojta, \Journal{\RPP}{66}{2069}{2003}.
\bibitem{Belitz05} D. Belitz, T.R. Kirkpatrick, T. Vojta, \Journal{\RMP}{77}{579}{2005}.
\bibitem{Cejnar08} P. Cejnar, J. Jolie, {\PPNP} (2008), in press; see arXiv:0807.3467[nucl-th].
%
\bibitem{Dieperink80} A.E.L. Dieperink, O. Scholten, F. Iachello, \Journal{\PRL}{44}{1747}{1980}.
\bibitem{Zhang87} W.-M. Zhang, D.H. Feng, J.N. Ginocchio, \Journal{\PRL}{59}{2032}{1987}.
\bibitem{Volya03} A. Volya, V. Zelevinsky, \Journal{\PLB}{574}{27}{2003}.
\bibitem{Rowe04} D.J. Rowe, \Journal{\PRL}{93}{122502}{2004}.
\bibitem{Emary03} C. Emary, T. Brandes, \Journal{\PRL}{90}{044101}{2003}.
%
\bibitem{Heinze06} S. Heinze, P. Cejnar, J. Jolie, M. Macek, \Journal{\PRC}{73}{014306}{2006}; M. Macek, P. Cejnar, J. Jolie, S. Heinze, \Journal{\PRC}{73}{014307}{2006}.
\bibitem{Cejnar06} P. Cejnar, M. Macek, S. Heinze, J. Jolie, J. Dobe{\v s}, \Journal{\JPAold}{39}{L515}{2006}. 
\bibitem{Cejnar07} P. Cejnar, S. Heinze, M. Macek, \Journal{\PRL}{99}{100601}{2007}.
\bibitem{Caprio08} M.A. Caprio, P. Cejnar, F. Iachello, \Journal{\ANNP}{323}{1106}{2008}.
%
\bibitem{Reichl} L.E. Reichl, {\em A Modern Course in Statistical Physics} (Wiley, New York, 1998).
\bibitem{Iachello06} F. Iachello, {\it Lie Algebras and Applications}, Lecture Notes in Physics, Vol.\,708 (Springer, Berlin, 2006).
\bibitem{Frank94} A. Frank, P. Van Isacker, {\it Algebraic Methods in Molecular \& Nuclear Structure Physics} (Willey, New York, 1994). 
\bibitem{Cejnar07b} P. Cejnar, F. Iachello, \Journal{\JPA}{40}{581}{2007}.
%
\bibitem{Stewart82} I. Stewart, \Journal{\RPP}{45}{185}{1982}.
\bibitem{Lipkin65} H.J. Lipkin, N. Meshkov, N. Glick, \Journal{\NPA}{62}{188, 199, 211}{1965}.
\bibitem{Vidal06} J. Vidal, J.M. Arias, J. Dukelsky, J.E. Garc{\'\i}a-Ramos, \Journal{\PRC}{73}{054305}{2006}.
\bibitem{Cary92} J.R. Cary, P. Rusu, \Journal{\PRA}{45}{8501}{1992}.
\bibitem{Leyvraz05} F. Leyvraz, W.D. Heiss, \Journal{\PRL}{95}{050402}{2005}.
\bibitem{Curro08} F. P{\'e}rez-Bernal, F. Iachello, \Journal{\PRA}{77}{032115}{2008}.
%
\bibitem{Bohr98} A. Bohr, B. Mottelson, {\em Nuclear Structure}, Vol.~2 (World Scientific, Singapore, 1998).
\bibitem{tobe} P. Str{\' a}nsk{\' y} {\it et al.}, in preparation.


\endthebibliography
\end{document}